\documentclass[twocolumn,preprintnumbers,superscriptaddress,amsmath,amssymb,longbibliography]{revtex4-2}

\usepackage{graphicx}
\usepackage{dcolumn}
\usepackage{float} 
\usepackage{bbm}
\usepackage[dvipsnames]{xcolor}
\usepackage[colorlinks,linkcolor=blue,urlcolor=blue,citecolor=blue,anchorcolor=blue]{hyperref}
\usepackage{setspace}
\usepackage{xpatch} 

\usepackage[dvipsnames]{xcolor}

\begin{document}
\title{Beyond Photon Shot Noise: Chemical Limits in Spectrophotometric Precision}

\author{Georg Engelhardt}
\email{georg-engelhardt-research@outlook.com}
\affiliation{ International Quantum Academy, Shenzhen 518048, China}

\author{Dahai He}
\affiliation{Department of Physics, Xiamen University, Xiamen 361005, China}

\author{JunYan Luo}
\affiliation{Department of Physics, Zhejiang University of Science and Technology, Hangzhou 310023, China}

\date{\today}

\pacs{
  }

\begin{abstract}
In this work, we investigate precision limitations in spectrophotometry (i.e., spectroscopic concentration measurements) imposed by chemical processes of molecules. Using the recently developed Photon-resolved Floquet theory, which generalizes Maxwell-Bloch theory for higher-order measurement statistics, we analyze a molecular model system subject to chemical reactions whose electronic and optical properties depend on the chemical state. Analysis of sensitivity bounds reveals: (i) Phase measurements are more sensitive than intensity measurements; (ii) Sensitivity exhibits three regimes: photon-shot-noise limited, chemically limited, and intermediate; (iii) Sensitivity shows a turnover as a function of reaction rate due to the interplay between coherent electronic dynamics and incoherent chemical dynamics. Our findings demonstrate that chemical properties must be considered to estimate ultimate precision limits in optical spectrophotometry.
\end{abstract}

\maketitle

\allowdisplaybreaks

\section{Introduction}

\begin{figure*}
	\includegraphics*[width = 1.\linewidth]{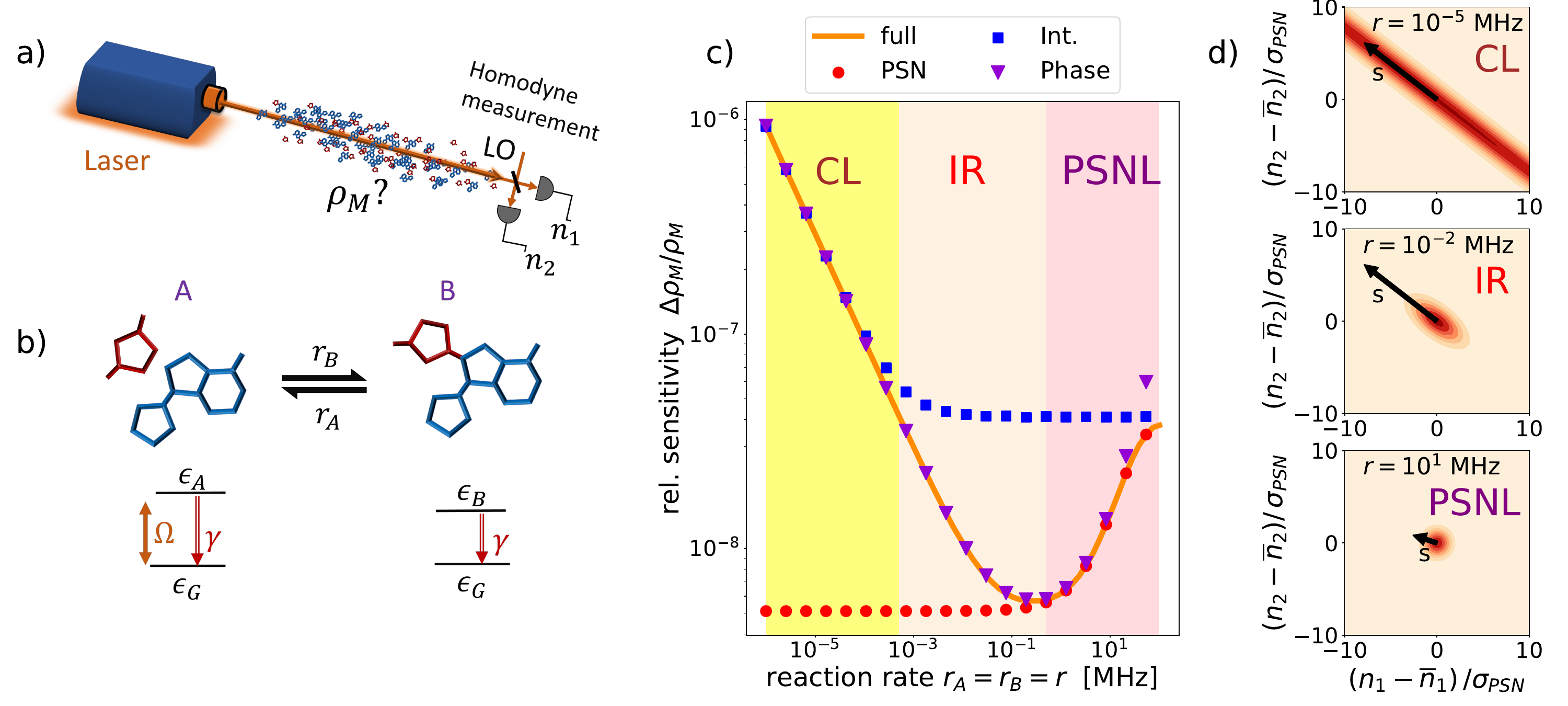}
	\caption{(a) Prototypical spectrophotometric measurement setup: coherent laser light propagates through a sample containing molecules with concentration $\rho_{M}$. The probe laser is measured by a homodyne setup using a local oscillator and two photon detectors. (b) Model molecule undergoing chemical reactions with rates $r_A$ and $r_B$, with electronic structure depending on the chemical state. (c) Relative sensitivity as a function of chemical reaction rate [Eq.~\eqref{eq:cramerRaoBound}, solid line]. Squares and triangles show sensitivity from intensity [Eq.~\eqref{eq:cramerRaoBoundIntensity}] and phase [Eq.~\eqref{eq:cramerRaoBoundPhase}] measurements; solid line shows the joint measurement sensitivity; circles show a naive photon-shot-noise estimate. The chemical-limited (CL) regime, the intermediate regime (IR), and the photon-shot-noise limited (PSNL) regimes are highlighted in colors. (d) Probability distributions in the three  sensitivity regimes. $\boldsymbol s =\partial_{\rho_M} \overline {\boldsymbol n}$ denotes the signal vector. Parameters are explained in Sec.~\ref{sec:molecularSystem}, and $\epsilon_{\Delta} =40\; \text{MHz}$.}
	\label{fig:overview}
\end{figure*}

Spectrophotometry---the measurement of light absorption to determine chemical concentrations—is as one of the most important analytical techniques across scientific disciplines, including medicine, environmental monitoring, biology and engineering~\cite{CorderoBorboa2020,Kazakov2021,Hurlbut1977,Haugen2026,Ciaffoni2013,ReyesReyes2015,Zhu2024,Ma2025a,Alden2023}. At its core lies Beer's law, relating absorption to concentration.
Practical spectrometers face a series of technical limitations, such as sample positioning uncertainties, optical reflections, electronic noise in detectors, and sample preparation inconsistencies~\cite{Soovaeli2006,Galban2007,Dobiliene2010,Peest2017,Skoog2017}.  Nevertheless, technological improvements in laser spectroscopy—both in intensity and phase measurement—continue to push the boundaries of classical performance~\cite{Demtroeder2014,Gazizov2025}. 

From a more fundamental point of view, quantum optics suggests photon-shot-noise, arising from the quantum nature of light itself, as the fundamental boundary setting the ultimate precision limitation~\cite{Escher2011,Taylor2016}. This perspective has fueled remarkable advances in quantum-enhanced spectroscopy, with squeezed and entangled light sources promising to  achieve unprecedented sensitivity~\cite{Li2022,Zhang2025,Dorfman2021,Zhang2025a,Zhang2022,Fan2024}.

In this paper, we argue that the fundamental precision limit in spectrophotometry lies  not only in the quantum statistics of light but also in the intrinsic chemical dynamics of the molecules being measured. While single-molecule spectroscopy has long revealed how chemical processes—spectral diffusion, conformational changes, chemical reactions—cause fluctuations in optical properties~\cite{Moerner2002,Mukamel2003,Barkai2004,Osadko2005}, these effects are often dismissed as irrelevant to ensemble measurements where averaging supposedly washes out individual molecular fluctuations. Sophisticated theoretical frameworks, from polaron transformations to hierarchical equations of motion, have been developed to understand these molecular dynamics~\cite{Silbey1984,Harris1985,Xu2016,Tanimura1989,Tanimura2020,Makri1995,Zhang2024}, but their implications for ensemble measurement precision remain largely unexplored.

To investigate this, we develop an operational framework that treats spectrophotometry as an estimation problem, where chemical dynamics directly influence measurement statistics. Our approach leverages the recently developed Photon-resolved Floquet Theory (PRFT)~\cite{Engelhardt2025}, which generalizes the Maxwell-Bloch theory to predict the complete measurement statistics of time-integrated intensities. This enables rigorous application of estimation theory—specifically the Cramér-Rao bound—to determine fundamental sensitivity limits. We analyze a prototypical molecular system undergoing chemical reactions between states with distinct optical properties, a model encompassing conformational changes, configurational transitions, and chemical reactions alike.

The model analysis yields conclusions representative for more sophisticated molecular systems:
\begin{itemize}
	
	\item Phase measurements  outperform intensity measurements, leveraging coherent quantum effects rather than incoherent absorption. This superiority persists across a wide parameter range.
	
	\item As chemical reaction rates vary, measurement sensitivity transitions through three distinct regimes:  (i) a \textit{photon-shot-noise limited regime} at fast rates where quantum optical limits prevail; (ii) a \textit{chemically-limited regime} at slow rates where molecular fluctuations dominate; and (iii) an \textit{intermediate regime} where intensity and phase measurements are limited by different mechanisms.
	
	\item Sensitivity does not monotonically improve with faster chemical rates. Instead, a turnover emerges where increasing rates initially improve precision by averaging molecular fluctuations, but eventually degrade it by destroying electronic coherence essential for an effective phase-sensitive measurement.
\end{itemize}

These findings  demonstrate that chemical properties are not merely sample characteristic to be measured, but actively influence the measurement precision itself. They reveal a rich interplay of chemical dynamics and quantum optics that sets the ultimate performance limits.

The remainder of this paper is structured as follows: Sec.~\ref{sec:systemAndMethods} introduces the molecular model and theoretical framework, including the PRFT methodology and sensitivity analysis. Sec.~\ref{sec:results} presents detailed results across the parameter space, explaining the three regimes and their physical origins. Sec.~\ref{sec:conclusions} summarizes our findings and discusses future research directions. Appendices provide technical details of the PRFT framework and analytical derivations.

\section{System and methods}

\label{sec:systemAndMethods}

\subsection{Molecular system }

\label{sec:molecularSystem}

The experimental setup of a common spectrophotometry device is sketched in Fig.~\ref{fig:overview}(a), in which a laser emits strong coherent light which propagates through a sample containing the molecules whose concentration $\rho_{M}$ is to be determined. After propagation, the light field is measured by a detector setup. Typically, the measurement is restricted to intensity measurements, but here we envision a homodyne detection, which allows access to both  the intensity and the phase of the probe light.

 The molecules exhibit a chemical reaction between two states  $A$ and $B$ with distinct electronic and optical properties.  The electronic structure of the molecule in each chemical state is sketched in Fig.~\ref{fig:overview}(b), which features distinct excitation energy $\epsilon_\alpha$ or transition dipole $d_\alpha$ depending on state $\alpha=A,B$. The reaction rates between these states are  $r_A$ and $r_B$.

 We deploy  the PRFT to calculate the probe light modification due to the interaction of the molecules~\cite{Engelhardt2025}. The molecular dynamics follow the semiclassical quantum master equation
\begin{eqnarray}
\frac{d}{dt}\rho &=&- \frac{i}{\hbar}\left[\hat H_0, \rho \right] +\mathcal L_{\text{dis}} \rho,
\label{eq:masterEquation:molecule}
\end{eqnarray}
in which the Hamiltonian $\hat H_0  = \hat H_{\boldsymbol \varphi = 0} $ is given by
\begin{equation}
	\hat H_{\boldsymbol \varphi}= \sum_{\alpha=A,B }\epsilon_\alpha \left|e_\alpha \right>\left< e_\alpha \right|  +  \hbar  \left( \frac{\Omega_{\alpha,\boldsymbol \varphi }}{2} \left|e_\alpha \right>\left< g_\alpha \right| +\text{h.c.} \right).
	\label{eq:Hamiltonian}
\end{equation}
Thereby,  $\left|g_\alpha \right> $  ($\left|e_\alpha \right> $) denote the ground (excited)  state of the molecule in chemical state $\alpha$. Probe-field induced  ground-to-exited-states transitions are described by  Rabi frequencies
$\Omega_{\alpha,\boldsymbol \varphi} = - \boldsymbol d_\alpha \cdot \boldsymbol E/\hbar  = \frac{ d_\alpha  E}{\sqrt{2} \hbar  }  \left( e^{i(\varphi_1 +\pi/4)} + e^{i(\varphi_2- \pi/4)}  \right) $,  proportional to the scalar product  of state-dependent transition dipoles $\boldsymbol d_\alpha$ and the probe laser electric field $\boldsymbol E $. The second equality expresses the Rabi frequencies via auxiliary phases $\varphi_k$ with $k=1,2$, allowing to define the photon-flux operators
 \begin{equation}
 \hat j_{k}  = \frac{1}{\hbar}\frac{d}{d\varphi_k} \hat 	H_{\boldsymbol \varphi} .
 \end{equation}
 According to the PRFT, these photon-flux operators  directly relate to the measurement statistics at  detector $k=1,2$ in Fig.~\ref{fig:overview}(a).
 
 The Liouvillian in Eq.~\eqref{eq:masterEquation:molecule} captures the dissipative dynamics of the molecule 
  \begin{eqnarray}
 \mathcal L_{\text{dis}} \rho &=& \gamma\sum_{\alpha=A,B} D \left[\left|g_\alpha \right>\left< e_\alpha \right|  \right]\rho \nonumber \\
 &+& \sum_{\alpha=A,B} r_\alpha \left( D \left[\left|g_{\alpha} \right>\left< g_{\overline\alpha} \right|  \right] \rho +    D\left[\left|e_{\alpha} \right>\left< e_{\overline\alpha} \right|    \right] \rho  \right)  ,
 \label{eq:dissipator}
 \end{eqnarray}
where $D[A]\rho = A\rho A^\dagger -\left\lbrace A^\dagger A,\rho \right\rbrace/2$ denotes the common dissipator. The first line describes spontaneous decay from the excited to the ground state with dissipation rate $\gamma$. The second line models the chemical reaction dynamics, assuming incoherent dynamics which does not distinquish ground and excited states for simplicity. The notation $\overline \alpha$ refers to $\overline A =B$ and $\overline B =A$.

In the numerical calculations we assume the following  realistic parameters: The probe laser has power $P = 1\,\text{mW} $, wavelength $\lambda = 500\, \text{nm}$, and diameter $d = 0.5\, \text{cm}$. The optical transition dipole of the molecules is $d_A = 1\,\text{Debye}$ and $d_{B}=0$. These parameters result in a Rabi frequency of $\Omega = 0.95\; \text{MHz}$.  The dissipation rate is assumed to be $\gamma = 10\;\text{MHz}$. The total measurement time is $\tau = 1\,\text{s}$.

\subsection{Measurement sensitivity}

The homodyne setup in Fig.~\ref{fig:overview}(a)  measures the probe field intensities incident to detector $k=1,2$ as a function of time. Analysis of this measurement data provides information about the light-matter interaction, such as the density of molecules. The most common observable deployed in this context is the time-integrated intensity operator defined as
\begin{equation}
		\hat n_{k} = \frac{\mathcal A}{\hbar  \omega_{\text{p}}} \int_{0}^{\tau} \hat I_{k} (t)dt,
\end{equation}
where $\hat I_{k} (t)$ is the intensity incident at detector $k$ (in the Heisenberg picture) and $\tau$ is the total measurement time. To represent this observable in units of photon numbers, we have multiplied it with the probe laser area $\mathcal A$ and divided it by the probe field photon energy $\hbar \omega_{\text{p}}$. The mean and the covariance matrix of this operator are defined as
\begin{equation}
	\overline n_{k}  =  \left< \hat n_{k}  \right>,
\end{equation}
and
\begin{equation}
\left[ \boldsymbol \Sigma^2\right]_{k,l}  =  \left< \hat n_{k} \hat n_{l}    \right>-\left< \hat n_{k}  \right>\left< \hat n_{l}  \right>,
\end{equation}
respectively. To estimate the sensitivity $\Delta X = \left<\Delta \hat X^2\right>^{1/2}$ of a system parameter $X$, we can take advantage of the celebrated Cram\'er-Rao bound~\cite{Cramer1946,Rao1945}, which for Gaussian observables reads
\begin{equation}
\left<\Delta \hat X^2\right>  =
\left[ \left( \partial_{X} \boldsymbol {\overline n}\right) \boldsymbol \Sigma ^{-2} \left( \partial_{X} \boldsymbol {\overline n}\right)^T \right]^{-1},
\label{eq:cramerRaoBound}
\end{equation}
and requires the evaluation of the means $\boldsymbol {\overline n}= (\overline n_1,\overline n_2)$  and covariance of the operators $\hat n_k$. For later purpose, we also introduce $\hat n_{\pm} = \hat n_1 \pm\hat n_2$.

After mixing the probe field with the coherent local oscillator field at the beam splitter [see Fig.~\ref{fig:overview}(a)], we find
\begin{eqnarray}
\left(
\begin{array}{c}
\overline n_{1} \\
\overline n_{2}
\end{array}
\right) 
=
 \overline n_{+}	\left(
\begin{array}{c}
\cos^2(\frac{\pi}{4}+ \frac{\overline\varphi - \overline\varphi_{\text{LO}}}{2}  ) \\
\sin^2(\frac{\pi}{4}+ \frac{\overline\varphi - \overline\varphi_{\text{LO}}}{2} )
\end{array}
\right) ,
\label{eq:meanPhotonNumbersAFOangle}
\end{eqnarray}
where $\overline n_{+}$ equals the sum of photons in the probe field  $\overline n_{\text{p}}$ (after the interaction with the molecules)  and the local oscillator $\overline n_{\text{LO} }$, and we assume that $\overline n_{\text{LO} } =  \overline n_{\text{p}} =\overline n_{+}/2$.
 The probe field (local oscillator) has mean  phase $\overline\varphi $ ($\overline\varphi_{\text{LO} } $).  Choosing the local oscillator phase such that $\overline n_{1}=\overline n_{2}$, we find
\begin{eqnarray}
\partial_{X} \boldsymbol {\overline n}^{\text{T}}
=
\overline n_{\text{p}} \frac{d\overline \varphi }{dX}	\left(
\begin{array}{c}
1 \\
-1
\end{array}
\right) 
+\frac{d\overline n_{\text{p}}  }{dX} 	\left(
\begin{array}{c}
1 \\
1
\end{array}
\right) .
\label{eq:signalVector}
\end{eqnarray}
Restricting the data analysis to either intensity or phase measurement, the corresponding sensitivity bounds are given by
\begin{subequations}
	\label{eq:cramerRaoBoundSimple}
\begin{eqnarray}
\left<\Delta X^2\right>_{\mathcal I}  &=&
 \left(\frac{d\overline n_{\text{p}}  }{dX} \right)^{-2} \Sigma_+^2 , \label{eq:cramerRaoBoundIntensity} \\
\left<\Delta X^2\right>_{\varphi}  &=& \left(\overline n_{\text{p}} \frac{d\overline \varphi  }{dX} \right)^{-2} \Sigma_-^2,
\label{eq:cramerRaoBoundPhase}
\end{eqnarray}
\end{subequations}
 where $\Sigma_\pm^2 = v_\pm  \boldsymbol \Sigma^2 v_\pm^{\text{T}} $ with $v_+ =\left( 1,1 \right)$ and  $v_- =\left( 1,-1 \right)$ are the variance of the operators $\hat n_{\pm}$.

\subsection{Evaluation}

\begin{figure*}
	\includegraphics[width = 1.\linewidth]{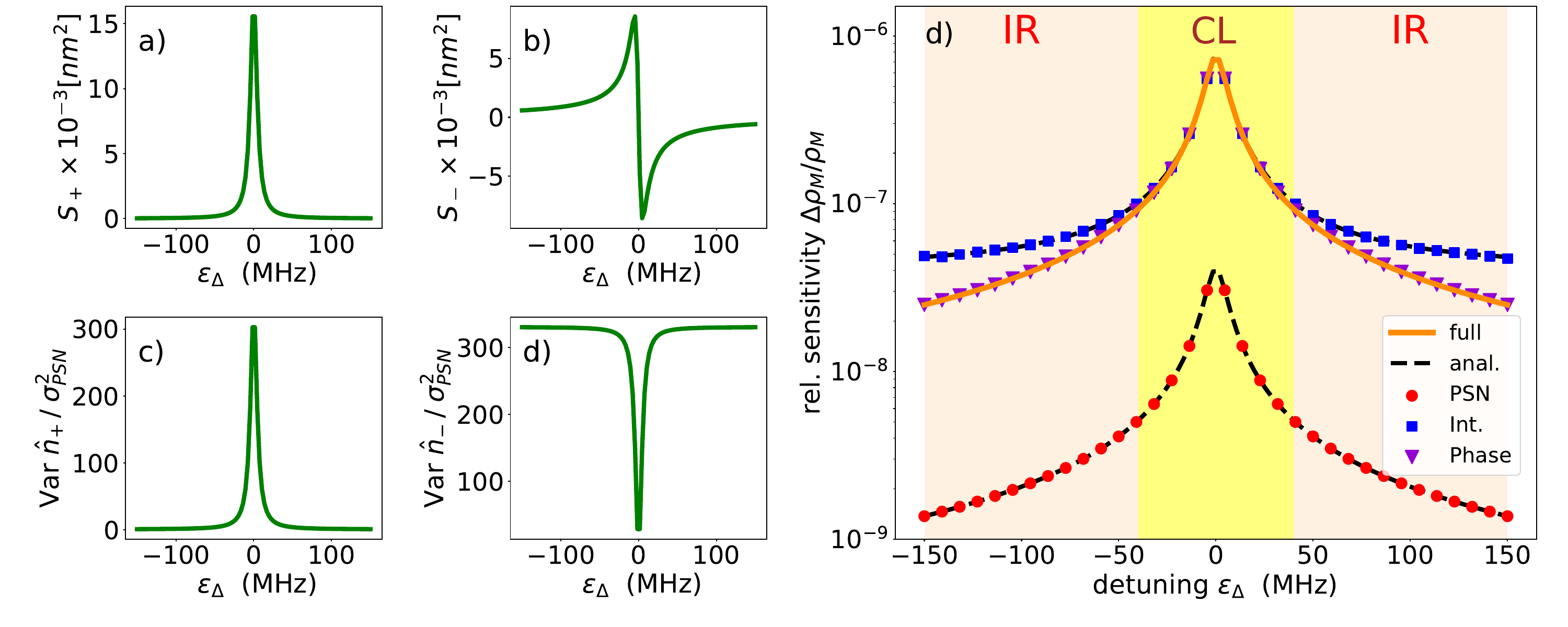}
	\caption{(a) Absorption cross section as function of detuning for the model system in Eq.~\eqref{eq:masterEquation:molecule} evaluated using Eq.~\eqref{eq:temporalAverage} for $\tau\rightarrow \infty$. (b) Phase-shift cross section evaluated using Eq.~\eqref{eq:temporalAverage}. (c) Variance of $\hat n_+$ of the time-integrated intensity measurements in Eq.~\eqref{eq:covariancePlusMinus}. (d) Variance of $\hat n_-$ calculated using the same equation. (e) Relative sensitivity as a function of detuning as predicted by full measurement statistics [Eq.~\eqref{eq:cramerRaoBound}, solid line], intensity measurement [squares, Eq.~\eqref{eq:cramerRaoBoundIntensity}], phase measurement [triangles, Eq.~\eqref{eq:cramerRaoBoundPhase}] and a simple photon-shot noise estimate (circles). Analytical calculations are depicted by a black dashed line. Overall parameters are explained in Sec.~\ref{sec:molecularSystem}, and $r_A =r_B =10^{-4} \text{MHz}$. }
	\label{fig:detuning}
\end{figure*}

To evaluate Eq.~\eqref{eq:cramerRaoBound}, we need to calculate the first two cumulants of the measurement statistics. We assume a sufficiently weak probe field ensuring the validity of Beer's absorption law,
\begin{eqnarray}
\overline n_{\text{p}}(z)   &=& \overline n_{\text{p},0}  e^{-\rho_{M}  \mathcal S_+  z},
\label{eq:beersLaw}
\end{eqnarray}
where $z$ is the propagation distance, $ \overline n_{\text{p},0} = \overline n_{\text{p}}(z=0)  $ is the initial photon number of the probe field, $\rho_{M}$ is the molecular density, and $ \mathcal S_+$ is the absorption cross section. The phase shift of the probe field fulfills
\begin{eqnarray}
\overline \varphi  
&=&  \rho_{M} \mathcal S_- z  ,
\label{eq:phaseShift}
\end{eqnarray}
where $\mathcal S_-$ is the phase-shift cross section. These cross sections fulfill 
 $\mathcal S_{\pm} = \mathcal S_{1}  \pm \mathcal S_{2} $ with
\begin{equation}
\mathcal S_k = \frac{1}{\mathcal J\tau}\int_{0}^{\tau} \left<  \hat j_{k}(t) \right>dt,
\label{eq:temporalAverage}
\end{equation}
where we have defined the photon-number intensity   $\mathcal J  =   \mathcal I/ \hbar \omega_{\text{p}}  = \overline n_{\text{p}} /( \mathcal A\tau)$ in terms of the common intensity $\mathcal I$. Relations \eqref{eq:beersLaw}-\eqref{eq:temporalAverage}  can be derived using the PRFT as shown in Appendix~\ref{app:photonResolvedFloquetTheory}.

Likewise,  the covariance matrix in the weak-probe-field regime is given by
\begin{equation}
 \boldsymbol \Sigma^2(z) 
= \boldsymbol \Sigma_0^2    e^{-2\rho_{M}  \mathcal S_+  z}   +  \int_0^{z} e^{-2\rho_{M}  \mathcal S_+  (z -z')} \boldsymbol  D_{\mathcal J (z') }dz'    ,
\label{eq:covarianceMatrix} 
\end{equation}
where $\boldsymbol \Sigma_0^2 = \overline n_{\text{p},0}\boldsymbol 1$ is the initial covariance matrix, and the diffusion matrix can be calculated via
\begin{equation}
\left[ \boldsymbol  D_{\mathcal J }  \right]_{k,l}  = \rho_{M}    \mathcal A \int_{0}^{\tau}\int_{0}^{\tau}  \left<\Delta \hat j_{k}(t_1) \Delta\hat j_{l}(t_2) \right>dt_1 dt_2,
\end{equation}
where $\Delta \hat j_{k}(t)  =  \hat j_{k}(t) -\langle\hat j_{k}(t) \rangle$. This relation is valid for a coherent local oscillator with $\overline n_{\text{LO} } =  \overline n_{\text{p}}(z)$ obeying Poisson statistics.

To make progress, we assume the following expansion of the diffusion matrix  in orders of the photon-number intensity
\begin{equation}
	 \boldsymbol  D_{\mathcal J}  = \rho_{M}   \tau \mathcal A \sum_{m=1}^{\infty} \frac{1}{m!} \boldsymbol  D^{(m)} \mathcal J^m .
	 \label{eq:diffusionMatrixExpansion}
\end{equation}
Our analysis for the molecular model in Eq.~\eqref{eq:masterEquation:molecule} will show that the diffusion matrix is fully characterized by the first two expansion  orders. Evaluating Eq.~\eqref{eq:covarianceMatrix}  using Eq.~\eqref{eq:diffusionMatrixExpansion}, we find
\begin{eqnarray}
\boldsymbol \Sigma^2 
&=&  \boldsymbol \Sigma_0^2   e^{-2\rho_{M}  \mathcal S_+  z}   \nonumber \\
 &+& \overline n_{\text{p},0} \frac{ \boldsymbol  D^{(1)}  }{\mathcal S_+}  \left( e^{-\rho_{M}  \mathcal S_+  z}   -  e^{-2\rho_{M}  \mathcal S_+  z}  \right)    \nonumber \\
&+& e^{-2\rho_{M}  \mathcal S_+  z}  \overline n_{\text{p},0} \mathcal J_0 \rho_{M} \boldsymbol  D^{(2)} z  ,
\label{eq:varianceFlow2} 
\end{eqnarray}
where $ \mathcal J_0 =\overline n_{\text{p},0}/(\mathcal A \tau) $and  we have neglected terms of order $\mathcal J^m$ with $m\ge 3$. 

In this work, we consider optical thick samples  such that the propagation length $z $ maximizes the signal intensity $\partial_{\rho_{M}} \overline n_{+}(z) $, that is, for $z=z_{\text{opt}} = 1/ \left( \rho_{M} \mathcal S_{+}\right)$. In doing so, we find that
\begin{subequations}
\begin{eqnarray}
\frac{\partial \overline n_{p}}{\partial \rho_{M}}   &=&  \frac{ \overline n_{\text{p},0}  }  {e \rho_{M} } ,
\qquad \frac{\partial \overline \varphi }{\partial \rho_{M}} 
= \frac{  \mathcal S_- }{\rho_{M} \mathcal S_+}  \label{eq:signalPlusMinus} \\
 \Sigma_{\pm}^2 
&=&  \frac{  \Sigma_{\pm,0}^2  }{e^2} + \overline n_{\text{p,0}}\frac{   D^{(1)}_{\pm}  }{\mathcal S_+} \left( \frac{1}{e} - \frac{1}{e^2} \right)    + \overline n_{\text{p},0}  \frac{ \mathcal J^{(0)} D_{\pm}^{(2)}  }{e^2  \mathcal S_+},\nonumber \\
\label{eq:covariancePlusMinus}
\end{eqnarray}
\end{subequations}
which can be used to evaluate Eq.~\eqref{eq:cramerRaoBoundSimple}. Thereby, we also defined $D_{\pm  }    = v_\pm \boldsymbol  D^{(m)}     v_\pm^{\text{T}}    $.

\section{Results}

\label{sec:results}

We are now in position to analyze the molecular model in Eq.~\eqref{eq:masterEquation:molecule}. For simplicity,  we assume that the transition dipole matrix elements of state $B$ are  zero. Figure~\ref{fig:detuning}(a) and (b) depict the  absorption and phase-shift cross sections as a function of the probe-field detuning, respectively. The absorption cross section displays the common absorption dip at zero detuning, while the phase-shift cross section exhibits the well-known dispersive curve. As shown in Appendix~\ref{app:spectroscopicProperties}, in an adiabatic regime for which $\gamma \ll r_{A},r_{B}$, the cross sections are given by
\begin{eqnarray}
\mathcal S_{\pm}  
&=&  p_A   \mathcal S_{\pm\mid A} + p_B   \mathcal S_{\pm\mid B},
\label{eq:effectiveCrossSection}
\end{eqnarray}
where $p_{\alpha}  = r_{\alpha}/(r_A +r_B)$ is the stationary probability to be in the chemical state $\alpha$. The cross sections conditioned on state $\alpha =A$ are
\begin{eqnarray} 
\mathcal S_{+\mid A}  &=& \frac{1}{2}  \frac{\gamma \beta^2  }{ 4\epsilon_{\Delta}^{2} +\gamma^2  }   ,\nonumber \\  %
\mathcal S_{-\mid A} &=&    \frac{1}{2}  \frac{2\epsilon_{\Delta} \beta^2  }{4 \epsilon_{\Delta}^{2} +\gamma^2  } ,   %
\label{eq:lm-crossections}
\end{eqnarray}
with $\epsilon_{\Delta} =\epsilon/\hbar -\omega_{\text{p}}$ and $\beta = d_A \sqrt{2 \hbar \omega_{\text{p}}/ \epsilon_0 c}$ (speed of light $c$, dielectric constant $\epsilon_0$), while the cross sections $\mathcal S_{\pm \mid B} =0$ as $d_{B}=0$. Evidently, a finite $d_{B}$ would give rise to a double peak structure of the cross sections according to Eq.~\eqref{eq:effectiveCrossSection}. 

An example for the variances  of the observables $\hat n_+$ and $\hat n_{-}$ is depicted in  Fig.~\ref{fig:detuning}(c) and (d), respectively, in units of the $\sigma_{\text{PSN} }^2  \equiv 2 \overline n_{\text{p}}(z_{\text{opt}})$, which can be regarded as a naive photon-shot noise estimate neglecting the noise added by the light-matter interaction. We observe in  Fig.~\ref{fig:detuning}(c) and (d) that  the full variance can drastically exceed this simple shot-noise estimate by more than two orders of magnitude, emphasizing the importance to take the noise added by the light-matter interaction into account. Moreover, we also observe that the variance of $\hat n_+$ and $\hat n_{-}$ behave complementary, that is, the   $\hat n_+$ variance is large close to resonance, while  the  $\hat n_-$ variance is large for off-resonance detunings, which will be corroborated with analytical calculations below.

In the adiabatic regime, the matrix elements of the diffusion matrix are given by
\begin{eqnarray}
[ \boldsymbol D_{\mathcal J }]_{k,l} 
&=&   p_A  D_{k,l \mid A} + p_B  D_{k,l \mid B} \nonumber \\
&+& \rho_{M} \tau \mathcal A \mathcal J^2 t_R p_A p_B  \left( \mathcal S_{k \mid A}  - \mathcal S_{k  \mid B}  \right) \left( \mathcal S_{l \mid A}  - \mathcal S_{l  \mid B}  \right) , \nonumber \\
\label{eq:kappa2:adiabatic}
\end{eqnarray}
as shown in Appendix~\ref{app:spectroscopicProperties}, where we have defined the effective reaction time as
\begin{equation}
t_R  = \frac{ 1}{  r_{A} +r_{B}}.
\label{eq:effectiveReactionTime}
\end{equation}
The  first line terms of Eq.~\eqref{eq:kappa2:adiabatic} represent diffusion matrices conditioned on the state $\alpha =A,B$ weighted by the stationary state probabilities $p_\alpha$. This term dominates for small $t_R$. The simplicity of the molecular model allows for an evaluation of the terms $ D_{k,l \mid A}$, which are explicitly given in Eq.~\eqref{eq:diffusionMatrix:conditioned}. Moreover, the adiabatic diffusion terms $D_{\pm  }^{(m)}  $ related to intensity and phase measurements are given by
\begin{eqnarray} 
D_{+ \mid A}^{(1)} &=& D_{- \mid A}^{(1)} =    \frac{\gamma\beta^2}{ 4\epsilon_{\Delta}^2 + \gamma^2 } , \nonumber \\  
D_{+ \mid A}^{(2)} &=& \frac{\beta^4\gamma \left( 8\epsilon_{\Delta}^2 -6 \gamma^{2} \right) }{  \left(4\epsilon_{\Delta}^2  + \gamma^2 \right)^3},\nonumber \\  %
D_{- \mid A}^{(2)}  &=&   \frac{2\beta^4    }{\gamma \left(4\epsilon_{\Delta}^2  + \gamma^2 \right)} 
 - \frac{8 \epsilon_{\Delta}^2  \left(  4\epsilon_{\Delta}^{2} + 5 \gamma^{2} \right)\beta^4 }{\gamma \left(4\epsilon_{\Delta}^2  + \gamma^2 \right)^3}  .
\label{eq:tavisCummingsModel:aptitudes}
\end{eqnarray}
However,  numerical analysis shows that these terms hardly contribute to the variance as compared to the photon-shot-noise $\sigma_{\text{PSN} }^2$. Thus we focus on the term in the second line of Eq.~\eqref{eq:kappa2:adiabatic} in the following. We analyze the measurement statistics in three regimes, namely a photon-shot-noise limited regime for small $t_R$, a chemically-limited regime for large $t_R$, and an intermediate regime for modest $t_R$.

\subsection{Photon-shot-noise-limited regime}

In the photon-shot-noise-limited (PSNL) regime, the measurement variance is dominated by the first two lines in Eq.~\eqref{eq:varianceFlow2}. Comparison of Eqs.~\eqref{eq:lm-crossections} and \eqref{eq:tavisCummingsModel:aptitudes} reveals that $D^{(1)}_{\pm}  =2 \mathcal S_+$. For this reason, we directly find that
\begin{eqnarray}
\frac{\Sigma_{\pm}^2  }{\sigma_{\text{PSN} }^2 }  &\approx&  1,
\label{eq:noise:psnlr}
\end{eqnarray}
which means that the measurement statistics is given by the photon-shot noise.

The measurement sensitivity in the PSNL regime is highlighted in Fig. \ref{fig:overview}(b) in the large reaction rate regime, where the sensitivity estimated by the full Cram\'er-Rao bound in Eq.~\eqref{eq:cramerRaoBound} (solid line) coincides with the photon-shot noise estimate (circles). This agreement shows that noise added by the light-matter interaction has a relatively minor impact. For relatively small $r_A =r_B=r$ within this regime, the measurement sensitivity is dominated by the phase measurement, which exceeds the intensity measurement due to the favorable scaling   of $\mathcal S_-$ with $\epsilon_{\Delta}$ in comparison to $\mathcal S_+$. This scaling is a result of the coherent nature of the phase shift, which contrast the incoherent dynamics characterizing the absorption. Only for very large   $r\approx 10^2\; \text{MHz}$ does the full measurement sensitivity comes from the intensity measurement. Here, rapid incoherent transitions between the chemical states destroy any electronic coherence in  $A$, which is necessary to produce a  phase shift, but  do not influence the absorption.

Because of the  Poisson statistics of the initial probe light,  the relative concentration measurement  sensitivity  is ultimately limited by
\begin{eqnarray}
\frac{\Delta \rho_{M} }{\rho_{M}}  &\rightarrow&\sqrt{ \frac{ e }{\overline n_{\text{p},0} } }
\end{eqnarray}
in the $r\rightarrow \infty$ limit, obtained by combining Eqs.~\eqref{eq:cramerRaoBoundSimple}, \eqref{eq:signalPlusMinus},  and \eqref{eq:noise:psnlr} for the intensity measurement. Thus, the relative sensitivity is only limited by the total number of photons deployed during the measurement. We remark that this does not imply that the measurement sensitivity can be arbitrarily improved by a large probe laser intensity, as our analysis is restricted to weak probe fields. However, as $\overline n_{\text{p}}^{(0)} \propto \tau$,  the measurement precision improves with the square root of the measurement time.

We depict the probability distribution in the PSNL regime  as a function of the photon numbers $n_k$  measured at detector $k$ in the lower panel of Fig.~\ref{fig:overview}(d), in which the probability distribution is equal along the directions $n_+ = n_1 +n_2$ and $n_- = n_1 - n_2$. Despite the relatively small noise, the sensitivity is  low, as the signal vector $ \boldsymbol{s}  = \partial_{\rho_{M}} \overline{\boldsymbol n}$ is small, which is a consequence of the rapid chemical transitions destroying electronic coherence in state $A$. 

\subsection{Chemically-limited regime}

For small  $r_A ,r_B$, inspection of Eq.~\eqref{eq:varianceFlow2} and Eq.~\eqref{eq:kappa2:adiabatic}  reveals that the measurement variance is dominated  by the second term proportional to $t_R$  in Eq.~\eqref{eq:effectiveReactionTime}. Assuming that $d_B =0$, we find  
\begin{eqnarray}
\frac{\Sigma_{\pm}^2  }{\sigma_{\text{PSN} }^2 }  &\rightarrow&  \frac{\overline n_{\text{p},0}}{e}\frac{p_B}{p_A} \frac{  \mathcal S_{\pm \mid A}^2 }{ \mathcal S_{+\mid A} \mathcal A} \frac{t_R}{\tau},
\label{eq:variance:chemicalLimited}
\end{eqnarray}
expressing the  variance  in terms of the cross sections $\mathcal S_{\pm \mid A}$. We  find that this term is suppressed for $p_A \gg p_B$, that is, for $r_A \gg r_B$. This defines a parameter regime, in which  the sensitivity is chemically limited (CL), which is marked in Fig.~\ref{fig:overview}(c).

Inserting the cross-section expressions of Eq.~\eqref{eq:lm-crossections} into Eq.~\eqref{eq:variance:chemicalLimited} yields
\begin{eqnarray}
\frac{\Sigma_{+}^2  }{\sigma_{\text{PSN} }^2 }  &\rightarrow&  \frac{p_B}{p_A} \frac{\gamma \Omega^2  }{ 4\epsilon_{\Delta}^{2} +\gamma^2  } \frac{t_R}{e} ,\nonumber \\
\frac{\Sigma_{-}^2  }{\sigma_{\text{PSN} }^2 }  &\rightarrow&  \frac{p_B}{p_A} \frac{4 \epsilon_{\Delta}^2 \Omega^2  }{\gamma \left(  4\epsilon_{\Delta}^{2} +\gamma^2  \right)} \frac{t_R}{e}
\label{eq:variance:intensityPhase}
\end{eqnarray}
for  the intensity and phase measurement variants, respectively, which reflects the functional form  in  Fig.~\ref{fig:detuning}(c) and (d). For an absorption measurement with $\epsilon_{\Delta} = 0 $, and for a phase measurement with $\epsilon_{\Delta} \gg \gamma$,  both cases scale $\frac{\Sigma_{\pm}^2  }{\sigma_{\text{PSN} }^2 } \propto \Omega^2 t_R /\gamma$, which allows to estimate the relative contribution of the chemical uncertainty to the total measurement noise.

Interestingly, using either expression in Eq.~\eqref{eq:variance:intensityPhase} to evaluate the  concentration measurement sensitivity Eq.~\eqref{eq:cramerRaoBoundSimple}, we find 
\begin{eqnarray}
\frac{\Delta \rho_{M} }{\rho_{M}}  &\rightarrow&\sqrt{ \frac{ p_B}{p_A} \frac{\mathcal S_+}{ \mathcal A }\frac{t_R}{\tau}},
\end{eqnarray}
which  depends on the ratio of the stationary probabilities, $\frac{ p_B}{p_A} $, the ratio of the absorption  and  probe field laser cross section, as well as the ratio of the effective chemical reaction time and the total measurement time.

The corresponding probability distribution is depicted in the upper panel Fig.~\ref{fig:overview}(d). While not fully aligned with the direction $n_-$, the probability distribution is more spread  in the direction $n_{-}$ than in the direction $n_+$, as for this off-resonance sensing the phase shift has a more substantial impact on the probe field than the absorption. The signal vector $\boldsymbol s$ is parallel to the long axes of the probability ellipsoid, which results from the quadratic dependence of the diffusion matrix in Eq.~\eqref{eq:kappa2:adiabatic} on the cross sections $\mathcal S_{k\mid \alpha}$ for large $t_R$.

\subsection{Intermediate regime}

Between the PSNL regime for short $t_R$ and the CL regime for long $t_R$, there is an intermediate regime (IR), which is marked in Fig.~\ref{fig:overview}(c). Here,  the intensity-measurement variance is dominated by the photon-shot noise, while the phase-measurement variance is dominated by the chemically-induced measurement noise. As  the phase-measurement is overall more sensitive than the absorption measurement, the full measurement sensitivity coincides with the sensitivity of the phase measurement.

The probability distribution in the intermediate regime is depicted in the middle panel Fig.~\ref{fig:overview}(d). While the distribution clearly expands in the $n_-$ direction related to the phase measurement, the extension in the $n_+$ hardly exceeds the photon-shot noise contribution (see bottom panel for comparison), reflecting that the absorption measurement is limited by the photon-shot noise, while the phase measurement is limited by the chemical noise.

\subsection{Detuning dependence}

\begin{figure*}
	\includegraphics[width = 1.\linewidth]{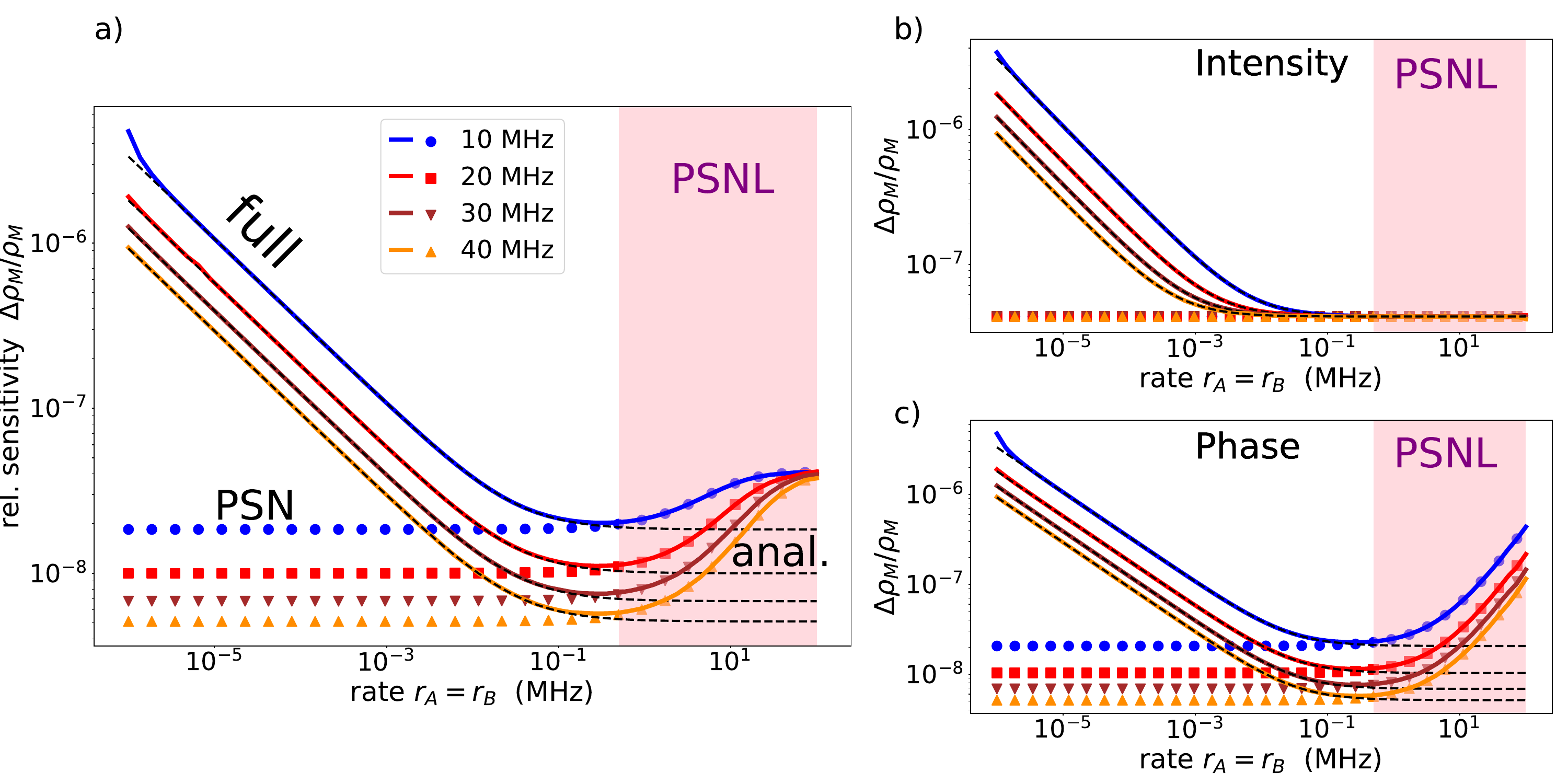}
	\caption{Relative sensitivity as a function of reaction rate evaluated using Eqs.~\eqref{eq:cramerRaoBound}, \eqref{eq:varianceFlow2}, and \eqref{eq:signalPlusMinus}. Solid lines depict sensitivity bounds determined by the full measurement statistics for various detunings, while markers show the sensitivity bounds predicted by the photon-shot-noise estimate. Dashed lines show the analytical calculations. (b) and (c) depict the same as (a), but restricted to  intensity and phase measurements, respectively. Overall parameters are explained in Sec.~\ref{sec:molecularSystem}.  }
	\label{fig:rate}
\end{figure*}

As the detuning is a crucial  control parameter, we discuss here in detail the sensitivity dependence on the detuning, which is depicted in Fig.~\ref{fig:detuning}(d) for a small $r_A = r_B = 10^{-4}\, \text{MHz} $. The full sensitivity [Eq.~\eqref{eq:cramerRaoBound}] is depicted by an orange solid line, while the sensitivity for the intensity measurement [Eq.~\eqref{eq:cramerRaoBoundIntensity}], the phase measurement [Eq.~\eqref{eq:cramerRaoBoundPhase}] and for the photo-shot-noise estimate are depicted by markers.  All sensitivity estimates  peak at $\epsilon_\Delta =0$, and then improve monotonically  with increasing detuning. Thereby, we can identify  two of the three sensitivity regimes:

For small detuning, we find that  the full sensitivity,  intensity-measurement sensitivity, and the phase-measurement sensitivity agree with each other, and fall short of the photon-shot-noise estimate. Consequently, the system is in the chemically-limited regime.   For large detuning $\left| \epsilon_{\Delta}\right| > 50\, \text{MHz}$, the sensitivity offered by the intensity measurement is larger than  the phase-measurement sensitivity, which still agrees with the full sensitivity. In this regime, the intensity-measurement noise is dominated by the photon-shot noise, such that the system is in the intermediate regime.

To further investigate the interplay between rate and detuning, we depict the relative sensitivity as a function of reaction rate for various detunings in Fig.~\ref{fig:rate}(a). For comparison, we added the sensitivity predicted by the photon-shot-noise estimate by markers in the same colors.   For all detunings, we observe a turnover as a function of the reaction rate, that is, an improving sensitivity for small reaction rates, and a deteriorating sensitivity for large reaction rates. As in the latter case the photon-shot-noise estimate agrees with the full sensitivity, we conclude that the system is in the PSNL regime.

To better understand the contributions of the intensity and the phase measurement to the full sensitivity, we depict the respective sensitivity estimates in Figs.~\ref{fig:rate}(b) and (c), respectively. The intensity-measurement sensitivity exhibits a monotonic decay for all detuning values. Beyond a detuning-specific reaction rate, the sensitivity saturates and agrees with the sensitivity predicted by the photon-shot-noise estimate. As this specific value does depend on the detuning, we do not highlight the  intermediate regime in the panels of Fig.~\ref{fig:rate}. Comparison of panels (a), (b) and (c) reveals that the turnover in the full sensitivity is inherited by the turnover in the phase-measurement sensitivity. In contrast to the intensity-measurement, the phase-measurement allows for a clear identification of the onset of the photon-shot-noise limited regime, which is independent of the detuning.

\section{Conclusions}

\label{sec:conclusions}
This work reveals that the ultimate precision of optical spectrophotometry is fundamentally constrained not only by the quantum statistics of light but also by the intrinsic chemical dynamics of the target molecules. By treating spectrophotometry as an estimation problem within the  framework of the Photon-resolved Floquet Theory (PRFT), we have quantified how chemical processes directly shape measurement statistics and set ultimate sensitivity bounds.

 First, phase-sensitive measurements consistently outperform intensity measurements by leveraging coherent quantum effects. Second, the analysis  identifies three distinct sensitivity regimes--- (i) a regime in which the sensitivity is limited by the photon-shot noise for large reaction rates; (ii) a regime in which the sensitivity is limited by chemical processes for  small reaction rates;  (iii) an intermediate regime in which the intensity measurement is limited by  photon-shot noise, while the phase-measurement is limited by chemical processes. Third, the interplay between coherent and incoherent dynamics gives rise to a turnover in the sensitivity as a function of the reaction rate. Our analysis demonstrates that the ultimate limit in spectroscopic concentration measurements is  set by an interplay between quantum optics and chemical dynamics. Ignoring the noise introduced by the light-matter interaction leads to an overly optimistic estimate of the achievable precision.

While our analysis using a simplified four-state model provides fundamental insights, it naturally opens avenues for future work. Incorporating coherent chemical transitions (e.g., via polaron-transformed master equation ~\cite{Wang2014,Wang2015,Xu2016,Liu2018,Dorfman2018a,Engelhardt2019},  HEOM methods~\cite{Tanimura2020}, or advanced quantum master equation approaches~\cite{Schaller2020,Becker2022}), photo-induced reaction pathways, and effects like rotational averaging or Doppler broadening into the PRFT framework will be essential for predicting precision limits in realistic, complex molecular systems. This work establishes a crucial step toward a complete quantum-mechanical theory of spectrophotometric precision, where the molecule is treated not as a static object but as a dynamic quantum entity whose fluctuations determine the ultimate measurement sensitivity.

\section*{Acknowledgments}

G.E. acknowledges  support by the National Natural Science Foundation of China (NSFC)  (Grant No. W2432004).   J.Y.L. acknowledges the support from the NSFC (Grant No. 11774311). D.H. acknowledges financial support from the NSFC 
(Grants No. 12475039) and the Guangdong Basic and Applied Basic Research Foundation (Grant No. 2025A1515010350).

\appendix

\section{Photon-resolved Floquet theory}

\label{app:photonResolvedFloquetTheory}

The PRFT adaptation to spectroscopy can be regarded as an extension of the celebrated Maxwell-Bloch theory capable of predicting the measurement statistics of time-integrated intensity measurements~\cite{Engelhardt2025}. Similar to the Maxwell-Bloch theory, the PRFT relies on the solution of semiclassical equations of motion of the matter system to retrieve the relevant information about the photon statistics of an electromagnetic probe field. Here, we consider the system in Sec.~\ref{sec:molecularSystem} assuming a probe field propagating through an ensemble of molecules and a homodyne measurement with a coherent local oscillator fulfilling $\overline n_{\text{LO} } =  \overline n_{\text{p}}(z)$. To obtain the photon statistics at the photon detectors $k=1,2$, we solve a two-sided quantum master equation of the form
\begin{eqnarray}
\frac{d}{dt}\rho_{\boldsymbol  \chi}= -i\left[H_{\boldsymbol \varphi+\frac{\boldsymbol  \chi}{2} } \rho_{\boldsymbol  \chi} -  \rho_{\boldsymbol  \chi} H_{\boldsymbol \varphi-\frac{\boldsymbol  \chi}{2} }  \right] +\mathcal L_{\text{dis}} \rho_{\boldsymbol  \chi},
\label{eq:twoSidedMasterEquation}
\end{eqnarray}
where the Hamiltonian is defined in Eq.~\eqref{eq:Hamiltonian} and depends on the so-called counting fields $\boldsymbol \chi =(\chi_1 ,\chi_2) $. Using the time-integrated and counting-field dependent density matrix, we define  the cumulant-generating function via
\begin{equation}
	\mathcal K_{ \boldsymbol  \chi  }  =\log \text{tr} \left[ \rho_{\boldsymbol  \chi} \right] .
\end{equation}
 In terms of the cumulant-generating function, we can construct flow equations for all relevant observable in spectroscopy, such as for the total photon number and the mean phase, which  read
\begin{eqnarray}
\frac{d{\overline n}_{+ } }{dz}  &=&\partial_z  {\overline n}_{ 2} + \partial_z {\overline  n}_{ 1},\nonumber \\\nonumber \\
\frac{d\overline \varphi }{dz}  
&=& \frac{\partial_z {\overline n}_{2} -\partial_z {\overline n}_{ 1}  } {\overline n_{+ }} . 
\label{eq:rotationAngel-photonNumber}
\end{eqnarray}
Thereby,  $\partial_z {\overline n}_{ k }  $ refers to the derivatives of the mean photon number ${\overline n}_{ k } $  with respect to the propagation distance $z$ and can be evaluated via
\begin{equation}
\partial_z {\overline n}_{ k }  =  \rho_{M}  \mathcal  A  \partial_{\chi_{k} }   \mathcal K_{\boldsymbol  \chi =0 } (z) .
\label{eq:fluxEvaluation}
\end{equation}
To evaluate Eqs.~\ref{eq:rotationAngel-photonNumber}, we further assume that  
\begin{equation}
	\partial_z {\overline n}_{ k }  = \rho_{M} \mathcal S_k {\overline n}_{+ }
	\label{eq:constantCrossSection}
\end{equation}
with a constant $\mathcal S_k$. This condition is typically fulfilled in the weak probe field regime. Using methods introduced in Ref.~\cite{Engelhardt2025} it is also possible to show that $ \mathcal S_k $ can be represented in the  integral form of Eq.~\eqref{eq:temporalAverage}. Integration of Eqs.~\eqref{eq:rotationAngel-photonNumber} now directly leads to Eqs.~\eqref{eq:beersLaw} and \eqref{eq:phaseShift}.

Likewise, one can construct a flow equation  for the covariance matrix related to time-integrated photon numbers, which reads
\begin{equation}
\frac{d}{dz}  \boldsymbol \Sigma^2 
=\boldsymbol  D_{\boldsymbol {\overline n} }  +   \boldsymbol   C_{\boldsymbol {\overline n} }  \boldsymbol   \Sigma^2   +\boldsymbol  \Sigma^2\boldsymbol   C_{\boldsymbol {\overline n}}  ,
\label{eq:varianceFlow} 
\end{equation}
and has to be solved with initial condition $\boldsymbol \Sigma^2 (0) = \overline n_{\text{p},0}\boldsymbol 1$. The coefficients can be expressed in terms of the cumulant-generating function as
\begin{eqnarray}
\left[	 \boldsymbol  D_{\boldsymbol {\overline n} }   \right]_{k,l}   &=&   \rho_{M}  \mathcal  A \;  \partial_{ \chi_{k} }\partial_{\chi_{l} }      \mathcal K_{ \boldsymbol \chi =0  } (z) \nonumber ,\\
\left[	 \boldsymbol  C_{\boldsymbol {\overline n} }  \right]_{k,l}   &=&  \rho_{M}  \mathcal  A\;   \partial_{\overline n_{k} }\partial_{\chi_{l} }   \mathcal K_{ \boldsymbol  \chi =0 } (z)-  \frac{1}{2}\frac{d\overline \varphi }{dz}  \hat \Xi \boldsymbol ,
\label{eq:diffusion-phaseSpaceMatrices}
\end{eqnarray}
where $\hat \Xi \boldsymbol =\hat \sigma_x + i\hat \sigma_y$. In this form, the flow equation describes the measurement statistics deploying a coherent local oscillator with Poisson photon statistics and $\overline n_{\text{LO} } =  \overline n_{\text{p}}(z)$, with its phase chosen such that  $\overline n_1 = \overline n_2$~\cite{Engelhardt2025}. Finally, under the assumption in Eq.~\eqref{eq:constantCrossSection}, one can show that 
\begin{equation}
	\boldsymbol  C_{\boldsymbol {\overline n} }   = \rho_{M} \mathcal S_+ \boldsymbol 1,
\end{equation}
which allows for an analytical integration of Eq.~\eqref{eq:covarianceMatrix}  yielding Eq.~\eqref{eq:varianceFlow2} .

\section{Spectroscopic properties of the molecule in the adiabatic regime}

\label{app:spectroscopicProperties}

\subsection{Adiabatic approximation}

In this appendix, we derive the expression for the spectroscopic properties of the molecular system in Sec.~\ref{sec:molecularSystem}, without  explicitly specifying the electronic properties in each state.  As the chemical transition described by the second line in Eq.~\eqref{eq:dissipator} is incoherent, we can neglect the corresponding coherence matrix elements in the density matrix and represent it as $ \boldsymbol \rho = \left( \rho_{A} ,\rho_B\right)  $. In this representation, the two-sided master equation in Eq.~\eqref{eq:twoSidedMasterEquation} reads
\begin{eqnarray}
\frac{d}{dt}  \boldsymbol \rho_{\boldsymbol  \chi}   = 
\left( 
\begin{array}{cc}
\mathcal L_{\boldsymbol  \chi \mid A }    &  \mathcal R_{AB} \newline \\
\mathcal R_{BA}       &   \mathcal L_{\boldsymbol  \chi \mid B} 
\end{array}
\right) 
\boldsymbol \rho_{\boldsymbol  \chi},
\label{eq:generalizedMasterEquation}
\end{eqnarray}
where $\mathcal L_{\boldsymbol \chi\mid \alpha} $ describes the dynamics within one of the chemical states $\alpha=A,B$, and $\mathcal R_{AB}, \mathcal R_{BA}$ represents the transition between the two chemical states.

Now we assume the following adiabatic ansatz for the counting-field dependent density matrix
\begin{equation}
\boldsymbol \rho_{\boldsymbol  \chi} = \left( p_{A ,\boldsymbol  \chi }\rho_{\boldsymbol  \chi \mid A}^{(0)} , p_{B,\boldsymbol  \chi} \rho_{\boldsymbol  \chi\mid B}^{(0)}\right),
\label{eq:ansatzDensityMatrix}
\end{equation}
where 
\begin{equation}
	\mathcal L_{\boldsymbol  \chi\mid \alpha} \rho_{\boldsymbol  \chi \mid \alpha}^{(0)}  = \mathcal K_{\boldsymbol  \chi \mid \alpha  }\rho_{\boldsymbol  \chi  \mid \alpha}^{(0)},
\end{equation}
such that $\mathcal K_{\boldsymbol  \chi \mid \alpha}$ is the eigenvalue of $\mathcal L_{\boldsymbol  \chi\mid \alpha} $ with the largest real part. Introducing the notation $ \boldsymbol p= \left( p_{A,\boldsymbol  \chi} ,p_{B,\boldsymbol  \chi }\right)  $, inserting Eq.~\eqref{eq:ansatzDensityMatrix} into the master equation in Eq.~\eqref{eq:generalizedMasterEquation}, and taking the partial trace over the electronic degrees of freedom, we obtain
\begin{eqnarray}
\frac{d}{dt}  \boldsymbol p   = 
\left( 
\begin{array}{cc}
\mathcal K_{\boldsymbol\chi \mid A}    &  r_{A} \newline \\
r_{B}       &   \mathcal K_{\boldsymbol \chi \mid B} 
\end{array}
\right) 
\boldsymbol p ,
\label{eq:effecitveMasterEquation}
\end{eqnarray}
where we have approximated  $\text{tr} \left [ \rho_{\boldsymbol\chi \mid \alpha}^{(0)}  \right]  \approx 1 $  and $ \text{tr} \left[ \mathcal R_{AB}  \rho_{\boldsymbol \chi\mid A}^{(0)}  \right] \approx r_A $, and $ \text{tr} \left[ \mathcal R_{BA}  \rho_{\boldsymbol\chi\mid B}^{(0)}  \right] \approx r_B$. This form is valid if $r_A,r_B  $ is smaller than the dissipative dynamics within each $\mathcal L_{\boldsymbol  \chi\mid \alpha } $, such that the stationary state of each chemical state $\alpha$ (represented by $\mathcal K_{\chi\mid \alpha}  $)  can be approached before the system undergoes a chemical transition. 
 
The statistics for long measurement times $\tau$ is determined by the dominating eigenvalue of the Liouvillian in Eq.~\eqref{eq:effecitveMasterEquation}, which explicitly reads
\begin{eqnarray}
\lambda_{\chi} &=& \frac{1}{2} \left(  \mathcal K_{\boldsymbol \chi \mid A} + \mathcal K_{\boldsymbol \chi \mid B} -r_{A}  -r_{B}  \right)   \nonumber\\
&+& \frac{1}{2}\sqrt{\left( \mathcal K_{\boldsymbol \chi \mid A}- \mathcal K_{\boldsymbol \chi \mid B} -r_{A}  +r_{B} \right)^2 +4  r_{A} r_{B}   } ,
\end{eqnarray}
such that 
\begin{equation}
 \mathcal K_{ \chi} = \lambda_{\chi} \tau
 \label{eq:cumulantGenFkt:adiabatic}
\end{equation}
is the asymptotic cumulant-generating function for $\tau \rightarrow \infty$, which can be used to evaluate~\eqref{eq:fluxEvaluation} and \eqref{eq:diffusion-phaseSpaceMatrices}.

Taking the first derivatives with respect to $\chi_k$ and using Eq.~\eqref{eq:effecitveMasterEquation}, we find 
\begin{eqnarray}
\partial_z {\overline n}_{ k }    
&=&   p_A   \partial_z {\overline n}_{k \mid A}   + p_B   \partial_z {\overline n}_{k\mid B}   ,
\label{eq:effectiveCrossSectionKappa}
\end{eqnarray}
where $\partial_z {\overline n}_{k \mid \alpha }  = -i \rho_{M} \mathcal A \partial_{\chi_k}\mathcal K_{ \chi =0 \mid \alpha} \tau $ is the photon-flux conditined on the chemcial state $\alpha$ and $p_\alpha = r_\alpha/(r_A +r_B)$ are the stationary probabilities. Using now Eqs.~\eqref{eq:fluxEvaluation} and \eqref{eq:constantCrossSection}, we finally arrive at Eq.~\eqref{eq:effectiveCrossSection}.

To evaluate the diffusion matrix in Eq.~\eqref{eq:diffusion-phaseSpaceMatrices}, we  derive Eq.~\eqref{eq:cumulantGenFkt:adiabatic} two times with respect to the counting fields and obtain
\begin{eqnarray}
D_{k,l} 
&=&   p_A  D_{k,l\mid A} + p_B  D_{k,l\mid B} \nonumber \\
&+&   \frac{ t_R p_A p_B }{ \rho_{M} \tau  \mathcal A }  \left( \partial_z {\overline n}_{k \mid A}   - \partial_z {\overline n}_{k \mid B}   \right) \left( \partial_z {\overline n}_{l\mid A}   - \partial_z {\overline n}_{l \mid B}   \right)  , \nonumber \\
\label{eq:diffusionMatrixElements}
\end{eqnarray}
where $ D_{k,l \mid \alpha } =  -  \rho_{M} \mathcal A \partial_{\chi_{k}  }\partial_{\chi_{l}  }\mathcal K_{ \chi =0 \mid \alpha}\tau$ are the diffusion matrix elements conditioned on the chemical state $\alpha$. Using again  Eq.~ \eqref{eq:constantCrossSection}, we eventually obtain Eq.~\eqref{eq:kappa2:adiabatic}.

\subsection{Diffusion matrix elements conditioned on the chemical state}

To fully evaluate Eq.~\eqref{eq:diffusionMatrixElements}, we must find a closed expression for the diffusion-matrix elements conditioned on the chemical state $\alpha$. As we assumed that $d_B=0$, we only have to evaluate the elements for $\alpha =A$. Therefore, we refer to Ref.~\cite{Engelhardt2025} which has derived these matrix elements already for a dissipative two-level system. In the weak-probe-field regime for which $\Omega^2/\gamma^2\ll 1$, the matrix elements become
\begin{eqnarray}
D_{k,l\mid A} &=&     \rho_{M} \mathcal A \left[   \frac{  a_0^{(k l)}  }{ a_1}  +  \frac{ 2 a_2 a_0^{(k) }  a_0^{(l) } }{ a_1^3}    \right. \nonumber \\
&-&  \left.   \frac{  a_0^{(k)}   a_1^{(l)}  + a_0^{(l)}  a_1^{(k)}  }{ a_1^2} \right]  ,
\label{eq:diffusionMatrix:conditioned}
\end{eqnarray}
where the coefficients are explicitly given by
\begin{eqnarray}
a_0^{(1)}&=& - i\left( \frac{\gamma}{8}      - \frac{\epsilon_{\Delta} }{4} \right) \Omega^2 , \nonumber \\
a_0^{(2)}&=&  - i\left( \frac{\gamma}{8}      + \frac{\epsilon_{\Delta} }{4} \right) \Omega^2  ,\nonumber \\
a_0^{(11)}&=& \frac{1}{8} \gamma \Omega^2 , \nonumber \\
a_0^{(22)}&=&  \frac{1}{8} \gamma \Omega^2  ,\nonumber \\
a_1&=&   \epsilon_{\Delta}^{2} + \frac{\gamma^2}{4},       \nonumber \\
a_1^{(1)}&=&  -i\frac{1}{4}  \Omega^2   , \nonumber \\
a_1^{(2)}&=& -i \frac{1}{4}  \Omega^2  , \nonumber \\
a_2&=&   \frac{\epsilon_{\Delta}^{2}}{\gamma} + \frac{5}{4} \gamma . 
\end{eqnarray}

\bibliography{projectLibrary}

\end{document}